\documentclass{aa}
\usepackage{psfig}
\psfull

\def\gsim{ \lower .75ex \hbox{$\sim$} \llap{\raise .27ex \hbox{$>$}} } 
\def\lsim{ \lower .75ex\hbox{$\sim$} \llap{\raise .27ex \hbox{$<$}} } 

\begin{document} 

\thesaurus{13(13.07.1; 02.16.2; 02.18.5)}

\title{GRB 990510: linearly polarized radiation from a 
fireball\thanks{Based on ESO VLT-Antu (UT1) observations (63.H-0233). 
Raw data are available upon request.}}

\author{Stefano Covino\inst{1} \and Davide Lazzati\inst{1,2} \and 
Gabriele Ghisellini\inst{1} \and Paolo Saracco\inst{1} \and Sergio Campana\inst{1}
\and Guido Chincarini\inst{1,2} \and Sperello Di Serego\inst{3} \and 
Andrea Cimatti\inst{3} \and Leonardo Vanzi\inst{4} \and Luca Pasquini\inst{4}
\and Francesco Haardt\inst{5} \and Gian Luca Israel\inst{6} \and Luigi Stella\inst{6}
\and Mario Vietri\inst{7}}

\institute{Osservatorio Astronomico di Brera, Via Bianchi 46, I--23807
Merate (Lc), Italy
\and
Dipartimento di Fisica, Universit\`a degli Studi di Milano,
Via Celoria 16, I--20133 Milano, Italy
\and
Osservatorio Astrofisico di Arcetri, 
Largo E. Fermi 5, I--50125 Firenze, Italy
\and
European Southern Observatory, Karl Schwarzschild Str. 2,
D--85748 Garching, Germany
\and
Universit\`a dell'Insubria, via Lucini 3, I--22100 
Como, Italy
\and
Osservatorio Astronomico di Roma, via Frascati 33,
I--00040 Monteporzio Catone, Roma, Italy
\and
Universit\`a di Roma III, via della Vasca Navale 84, 
I--00147 Roma, Italy
}

\offprints{S. Covino}
\mail{covino@merate.mi.astro.it}

\date{Received 1 June 1999; Accepted 14 June 1999}

\maketitle

\begin{abstract}
Models for gamma--ray burst afterglows envisage 
an hyper--relativistic fireball that is decelerated in the ambient 
medium around the explosion site. 
This interaction produces a shock wave which amplifies the magnetic field 
and accelerates electrons to relativistic energies, setting the  
conditions for an efficient production of synchrotron photons.
If produced in a region of large--scale ordered magnetic field, 
synchrotron radiation can be highly polarized.  
The optical transient associated with GRB~990510
was observed $\sim 18.5$ hr after the event and linear 
polarization in the $R$ band was measured at a level of $1.7\pm 0.2\%$. 
This is the first detection of linear polarization
in the optical afterglow of a gamma--ray burst.
We exclude that this polarization is due to dust in the interstellar
material, either in our Galaxy or in the host galaxy of the gamma--ray burst.
These results provide important new evidence in favor of  
the synchrotron origin of the afterglow emission,  
and constrains the geometry of the fireball and/or magnetic 
field lines.
\end{abstract}

\section{Introduction}
 
GRB 990510 was detected by BATSE on-board the Compton Gamma Ray Observatory
and by the {\it Beppo}SAX Gam\-ma Ray Burst Monitor and Wide Field Camera  
on 1999 May 10.36743 UT (\cite{kip99,ama99}, Dadina et al. 1999).
Its fluence (2.5$\times 10^{-5}$ erg cm$^{-2}$ above 20 keV) was 
relatively high (\cite{kip99}).
Follow up optical observations started $\sim 3.5$~hr later and revealed an
$R\simeq 17.5$ (\cite{axe99})
optical transient, OT (Vreeswijk et al. 1999a), at the coordinates (J2000)
$\alpha=13^{\rm h} 38^{\rm m} 07.11^{\rm s}$, $\delta=-80^\circ 29^\prime 
48.2"$ (\cite{hjo99b}) (galactic coordinates $\ell^{II}=304.942$, 
$b^{II}=-17.8035$).
Fig. 1 shows the Digital Sky Survey II image of the field of GRB 990510,
together with the European Southern Observatory (ESO) 
Very Large Telescope (VLT) image we obtained (see below): 
the OT is clearly visible in the latter. 

\begin{figure*}[!ht]
\centerline{\psfig{figure=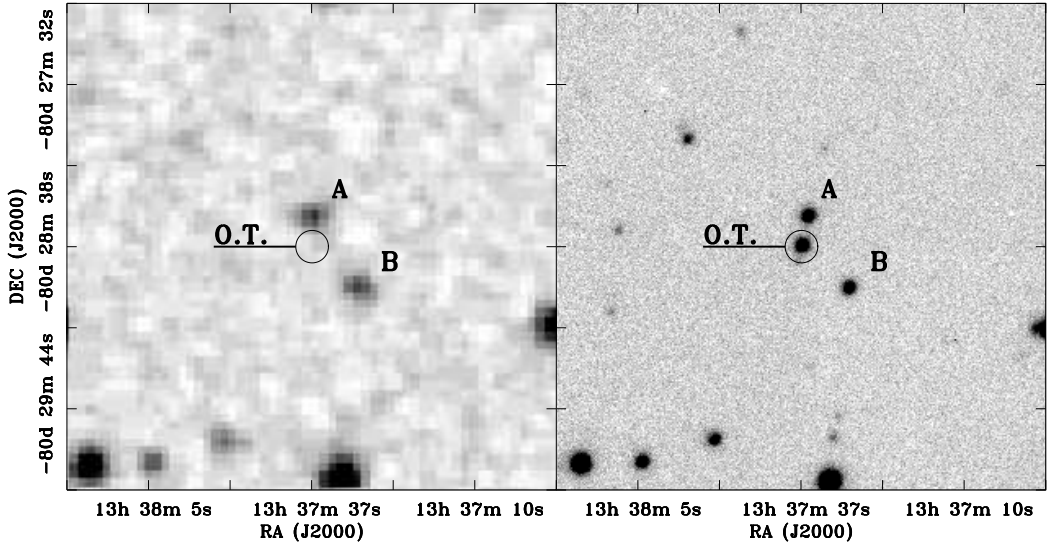,angle=90,width=14.0cm}}
\caption{{\it Left:} The Digital Sky Survey II image 
($1^\prime\times 1^\prime$) of the field of GRB 990510. 
The position of the optical transient is indicated by the circle. 
{\it Right:} The VLT image of the same field shows the optical transient at
the magnitude of $R\sim 19$, about 18 hours after the burst.
}
\end{figure*}

The OT showed initially a fairly slow flux decay $F_\nu\propto t^{-0.85}$
(Gala\-ma et al. 1999), which gradually steepened,
$F_\nu\propto t^{-1.3}$ after $\sim 1$~d (\cite{sta99a}),  
$F_\nu\propto t^{-1.8}$ after $\sim 4$~d (\cite{pie99}, Bloom et al 1999),
$F_\nu\propto t^{-2.5}$ after $\sim 5$~d (\cite{mar99}, 1999b)).
Vreeswijk et al. (1999b) detected Fe II and Mg II absorption lines in 
the optical spectrum of the afterglow.
This provides a lower limit of $z=1.619\pm 0.002$ to the redshift, and 
a $\gamma$--ray energy of $> 10^{53}$~erg, in the case of isotropic emission.

Polarization is one of the clearest signatures of synchrotron radiation, if 
this is produced by electrons gyrating in a magnetic field that is at least 
in part ordered.
Polarization measurements can provide a 
crucial test of the synchrotron shock model (\cite{mes97}).
An earlier attempt to measure the linear polarization of the optical
afterglow of GRB 990123 yielded only an upper limit (\cite{hjo99a})
of $\sim 2.3$\%. 

\section{Observations}

Our observations of GRB~990510 were obtained at ESO's VLT--Antu (UT1), equipped 
with the Focal Reducer/low dispersion Spectrometer (FORS) 
and Bessel filter $R$. 
The OT associated with GRB 990510 was observed $\sim 18.5$~hr 
after the burst, when the $R$-band magnitude was  $\sim 19.1$. 
Observations were performed in standard resolution
mode with a scale of $0.2\arcsec$/pixel; 
the seeing was $\sim 1.2\arcsec$. The observation log is reported in Table\,1.

\begin{table}
\begin{small}
\begin{center}
\begin{tabular}{cccc}
\hline
{\bf Starting time} & {\bf Exposure} & {\bf Angle} & {\bf filter} \\
{\bf UT, 1999/05/11}  & {\bf sec}      & {\bf degrees} & \\
\hline
02:48 & 600 & 00.0 & R \\
02:59 & 600 & 22.5 & R \\
03:10 & 600 & 45.0 & R \\
03:21 & 600 & 67.5 & R \\
\hline
\end{tabular}
\end{center}
\end{small}
\caption{Observation log for the polarimetric observation of the GRB\,990510 
field.}
\vskip -0.85truecm
\end{table}

Imaging polarimetry is achieved by the use of a Wollaston prism 
splitting the image of each object in the field
into the two orthogonal polarization components which appear in adjacent areas
of the CCD image.
For each position angle $\phi/2$ of the half--wave plate rotator,
we obtain two simultaneous images of cross--polarization, at angles 
$\phi$ and $\phi+90^\circ$. 

Relative photometry with respect to all the stars in the field 
was performed and each couple of simultaneous measurements at orthogonal angles 
was used to compute the points in Fig. 2 
(see Eq.~\ref{eq:sphi}). 
This technique removes any difference between the 
two optical paths (ordinary and extraordinary ray) and the polarization 
component introduced by  galactic interstellar grains along the line of sight. 
Moreover, being based on relative photometry in simultaneous images, 
our measurements are insensitive to intrinsic 
variations in the optical transient flux 
($\sim 0.03$ magnitudes during the time span of our observations).
With the same procedure, we observed also two polarimetric 
standard stars, BD--135073 and BD--125133, 
in order to fix the offset between the polarization and
the instrumental angles.

The data reduction was carried out with the ESO--MIDAS (version 97NOV) system.
After bias subtraction, non--uniformities were corrected using 
flat--fields obtained with the Wollaston prism. 
The flux of each point source
in the field of view was derived by means of both aperture and profile fitting
photometry by the DAOPHOT\,II package (\cite{ste87}), as implemented in MIDAS. 
For relatively isolated stars the two techniques differ only by 
a few parts in a thousand.

In order to evaluate the parameters describing the linear polarization of the 
objects, 
we compute, for each instrumental position angle $\phi$, the quantity:
\begin{equation}
S(\phi)\, =\, { \;\;
\frac{I(\phi)/I(\phi+90^\circ)}
{\langle I_{\rm u}(\phi)/I_{\rm u}(\phi+90^\circ)\rangle }-1 \;\;
\over
\frac{I(\phi)/I(\phi+90^\circ)}
{\langle I_{\rm u}(\phi)/I_{\rm u}(\phi+90^\circ)\rangle }+1
}
\label{eq:sphi}
\end{equation}
where $I(\phi)$ and $I(\phi+90^\circ)$ are the intensities of the object 
measured in the two beams produced by the Wollaston prism, and 
$\langle I_{\rm u}(\phi)/I_{\rm u}(\phi+90^\circ)\rangle $ are the average 
ratios of the intensities of the stars in the field. This corrects directly 
for the small instrumental polarization (and, at least in part, for the
possible interstellar polarization). 
These field stars (see Fig.\,3) have been selected over a 
range in magnitude ($18\le R\le 22$) to check for possible non--linearities. 
Since the interstellar polarization of any star in the field may be related
to the patchy dust structure and/or to the star distance, we have verified that 
the result does not depend on which stars are chosen for the analysis.
The parameter $S(\phi)$ is related to the degree of linear
polarization $P$ and to the position angle of the electric field vector 
$\vartheta$ by:
\begin{equation}
S(\phi)\, =\, P \cos 2(\vartheta - \phi).
\end{equation}
$P$ and $\vartheta$ are evaluated by fitting a cosine curve to the observed 
values of $S(\phi)$. 
The derived linear polarization 
of the OT of GRB 990510 is 
$P=(1.7\pm 0.2)$\% (1$\sigma$ error), at a position angle of  
$\vartheta=101^\circ \pm 3^\circ$\footnote{Please, note that the position angle
reported in IAUC\,7172 is incorrect by $90^\circ$}. The errors for the 
polarization level and position angle are computed propagating the photon 
noise of the observations and the contribution of the normalization to the 
stars in the field and of the calibration of the position angle. The latter 
quantities, however, amounts to only a minor fraction of the quoted 1$\sigma$ 
uncertainties. Fig. 2 shows the data points and the 
best fit $\cos\phi$ curve. 
The statistical significance of this measurement is very high. 
A potential problem is represented by a ``spurious'' polarization introduced 
by dust grains interposed along the line of sight, which may be preferentially 
aligned in one direction.
Stanek et al. (1999b), using dust infrared emission
maps (\cite{schle98}), reported a substantial Galactic absorption 
($E_{B-V}\simeq 0.20$) in the direction of GRB 990510. 
The maps by Dickey \& Lockman (1990)
and by Burstein \& Heiles (1982) give instead a somewhat lower value, 
$E_{B-V} \simeq 0.17$ and $\simeq 0.11$, respectively. 
Applying an empirical relation (\cite{hil56,ser75})
this polarization can amount
to $P_{\rm max} \le 9.0\,E_{\rm B-V}\,$, {\it i.e.} $\sim 1-2$\%. 
These are only statistical estimates and large variations on the main trend 
may be expected. 
\begin{figure}[!t]
\psfig{figure=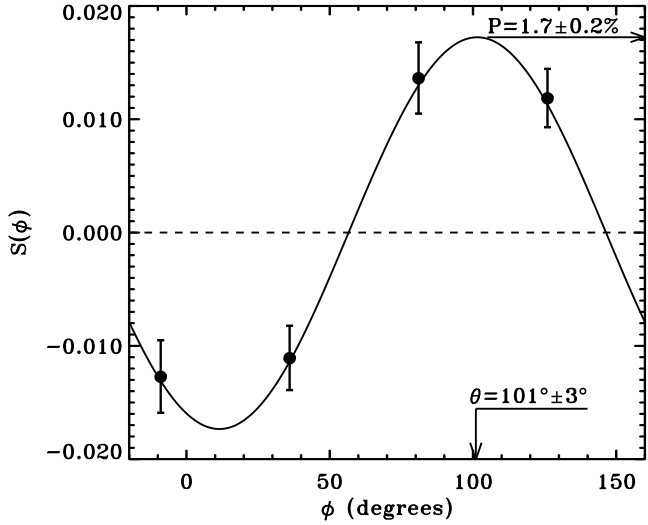,width=8.6cm}
\caption{Our polarization data taken at four different position angles
$\phi$ are fitted with a cosine curve.
The amplitude of this curve corresponds to the degree of linear polarization,
and its maximum to the polarization position angle.
Data are normalized to the average of the stars in the same field (see Eq. 1).
}
\end{figure}
Then a fraction, or even all the 
polarization of the OT could be 
caused by the passage of its light through the galactic ISM. 
However the normalization of the OT measurements to the stars 
in the field already corrects for the average interstellar 
polarization of these stars, 
even if this does not necessarily account 
for all the effects of the galactic ISM along the line of sight to the OT
(e.g. the ISM could be more distant than the stars, not
inducing any polarization of their light).
To check this possibility, we plot in Fig. 3 the degree of polarization
vs. the instrumental position angle for each star and for the OT.
All points in this figure have been derived avoiding to normalize with
respect to other objects.
It is apparent that, while the position angle of all stars are consistent 
with being the same (within 10 degrees), the OT clearly stands out.
The polarization position angle of stars close to the OT
differs by $\sim 45^\circ$ from the position angle of the OT (see Fig. 3).
This is contrary to what one would expect if the polarization of the OT 
were due to the galactic ISM. 
Indeed, the higher polarization level measured for the OT when 
normalized to the stars in the same field implies that 
the ISM actually {\it somewhat de-polarizes} the OT.
We therefore conclude that the OT, even if contaminated by interstellar 
polarization, must be intrinsically polarized to give the observed orientation.

\begin{figure}[!t]
\psfig{figure=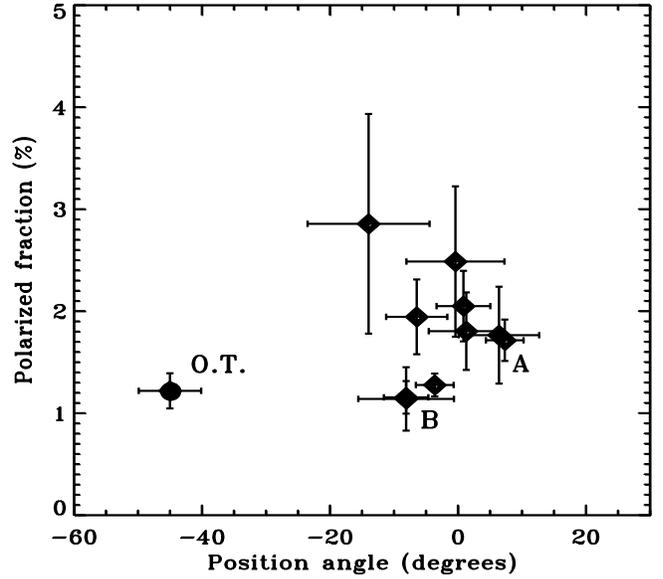,width=8.6cm,height=7.7cm}
\caption{The unnormalized 
degree of polarization vs the instrumental polarization 
position angle of the stars in the field and the optical transient.
The optical transient clearly stands out (P = $1.2 \pm 0.2$\%). Stars A and B
of Fig.\,1 are also labelled.}
\end{figure}

We can place tight limits on the amount of absorption, and hence the 
associated polarization, that could be produced by interstellar 
material in the host galaxy of GRB~990510. 
Assuming that the intrinsic spectrum is a power law 
$(F_\nu \propto \nu^{-\alpha})$, we require that the fluxes measured
simultaneously in the $B$, $V$, $R$ and $I$ band (Pietrzynski \& Udalski 1999; 
Kaluzny et al. 1999; Hjorth et al. 1999b) 
lie on a power law curve.
This strongly limits the amount of the local extinction, affecting
the flux at rest--frame frequencies of $\nu^{\prime} =(1+z)\nu_{obs}$,
i.e. in the UV, where extinction is more severe.
We find a maximum allowed value $E^{host}_{B-V}\sim 0.02$, corresponding
to a maximum induced polarization level of $\sim 0.2\%$.
Incidentally, the best fit power law is obtained for 
$\alpha\sim 0.7$,
a galactic $E_{B-V}=0.16$ and $E^{host}_{B-V}\sim 0$.
This value of $\alpha$ matches the predictions of the standard
model for the decaying afterglow flux (\cite{mes97}),
which gives $F_\nu(t)\propto t^{-3\alpha/2}$. 
For $\alpha=0.7$, the expected flux decay is in agreement with that 
measured at the time of the observations. 

\section{Discussion}

Relativistic fireball models do explain the main properties of GRBs 
and their afterglows (\cite{ree92,vie97,wax97,sar98}).
Polarized optical synchrotron emission may be observable 
if: (i) the coherence length of the magnetic field in the 
fireball grows at a sizeable fraction of the 
speed of light (Gruzinov \& Waxman 1999; Gruzinov 1999) or, (ii) the fireball 
is collimated (Hjorth et al. 1999a) (i.e. it is beamed). 
Therefore, measurements of optical polarization can provide constraints 
on the geometry of the emitting source. 

Additional information come from the afterglow light curve which shows
a gradual steepening in the bands $V$, $R$ and $I$, which was never observed 
before (Marconi et al. 1999a).
The observed steepening is almost wavelength independent, thus excluding 
that it could be entirely caused by a curved spectrum shifting 
in time rigidly to lower frequencies, in which case we ought to see 
the highest frequencies steepening first. 
In addition, the $V-R$ and $R-I$ colors are changing very slowly during 
the evolution, indicating that the spectral slope is changing slowly with time.  
These information suggest that the fireball is collimated in a jet.
The solid angle of the jet visible to the observer is limited to those 
regions making an angle smaller than $1/\Gamma$ with the line of sight.
As $\Gamma$ decreases, the visible solid angle increases as
$1/\Gamma^2$, until $\Gamma =\Gamma_1 = 1/(\theta_j-\theta)$
(with $\theta_j$ being the cone of semi--aperture angle and $\theta$ the angle 
between the cone axis and the line of sight).
When $\Gamma =\Gamma_2 = 1/(\theta_j+\theta)$ the observed solid angle 
remains constant, since the entire jet 
is visible.
For $\Gamma$--factors between $\Gamma_1$ and $\Gamma_2$
the observed solid angle increases
somewhat slower than $1/\Gamma^2$.
Since the flux at the earth is proportional to the observed solid angle,
we have two well defined behaviors of the light
curve, corresponding to $\Gamma>\Gamma_1$ and $\Gamma<\Gamma_2$, and
a transition period of gradual steepening in between. 
Photons produced in regions at an angle $1/\Gamma$ with respect to
the line of sight are emitted, in the comoving frame,
at $\sim 90^\circ$ from the velocity vector.
A comoving observer at this angle can then see a compressed emitting region 
and a projected magnetic field structure with a preferred orientation.
If the gradual steepening of the light curve is due to the 
mechanism just mentioned, 
we would observe only some regions at a viewing angle $1/\Gamma$, not all 
those we would see in axis--symmetric situation, and this asymmetry
can be the cause of the observed linear polarization.

The above arguments suggest that we are observing, slightly off--axis, 
a collimated beam.
If this is the case, we would have a link between the flux decay 
behavior, the presence of polarization, and the degree of collimation,
opening a new perspective for measuring the intrinsic power of GRBs.

Deeper understanding of polarization in GRBs may come from future 
multi--filter observations and from spectropolarimetry.
Frequency dependent polarization can in fact easily disentangle different
components of polarization.
In addition, variability in the degree of polarization and its position angle
is expected in such fastly evolving sources: therefore repeated
observations of the same afterglow will also be important.

\begin{acknowledgements}
We thank the ESO--VLT service team, and in particular 
H. Boehnhardt, F. Bresolin, P. M{\o}ller and G. Rupprecht.
FH thanks the kind hospitality of the Department of Physics of
the University of Milan. We also thank the referee J. Hjorth for his helpful
comments.
\end{acknowledgements}



\begin{thebibliography}{99}

\bibitem[Amati et al. 1999]{ama99} Amati, L., Frontera, F., Costa, E. \& Feroci, 
	M., 1999, GCN 317
\bibitem[Axelrod et al. 1999]{axe99} Axelrod, T., Mould, J \& Shmidt, B., 1999, GCN 315 
\bibitem[Bloom et al 1999]{blo99} Bloom, J. S., Kulkarni, S. R., Djorgovski, 
     S., Frail, D A., Axelrod, T.S., Mould, J.R., \& Shmidt, B.P., 1999, GCN 323
\bibitem[Burstein \& Heiles 1982]{bur82} Burstein, D., Heiles, C.,
	1987, AJ, 87, 1165
\bibitem[Dadina et al. 1999]{dad99} Dadina, M., Di Ciolo, L., Coletta, A., et al., 
     1999, IAUC 7160
\bibitem[Dickey \& Lockman 1990]{dic90} Dickey, J. M., Lockman, F.J., 1990,
	ARA\&A, 28, 215
\bibitem[Galama et al. 1999]{gal99} Galama, T. J., Vreeswijk, P.M., Rol, E. et al., 
	1999, GCN 313
\bibitem[Hiltner 1956]{hil56} Hiltner, W.A., 1956, ApJ, 2, 389
\bibitem[Hjorth et al. 1999a] {hjo99a} Hjorth, J. et al., 1999a,
	Science, 283, 2073
\bibitem[Hjorth et al. 1999b]{hjo99b}  Hjorth, J., Burud, I., Pizzella, A., Pedersen, H.,
	Jaunsen, A.O. \& Lindgren, B., 1999b, GCN 320
\bibitem[Kaluzny et al. 1999]{kal99} Kaluzny, J, Garnavich, P.M., Stanek, K.Z.,
	Pych W. \& Thompson, I., 1999, GCN 314
\bibitem[Kippen 1999]{kip99} Kippen, R. M., 1999, GCN 322
\bibitem[Gruzinov 1999]{gr99} Gruzinov, A., 1999, submitted to ApJ (astro-ph/9905276)
\bibitem[Gruzinov \& Waxman 1999]{gru99} Gruzinov, A. \& Waxman, E.,
	1999, ApJ, 511, 852
\bibitem[Marconi et al. 1999a]{mar99} Marconi, G., Israel, G.L., Lazzati, D., Covino, S. 
	\& Ghisellini, G., 1999a, GCN 329
\bibitem[Marconi et al. 1999b]{mar99b} Marconi, G., Israel, G.L., Lazzati, D., Covino, S. 
	\& Ghisellini, G., 1999b, GCN 332
\bibitem[M\'esz\'aros \& Rees 1997]{mes97} M\'esz\'aros,  P. \& Rees, M. J. , 1997,
	ApJ, 476, 232
\bibitem[Pietrzynski \& Udalski 1999]{pie99} Pietrzynski, G. \&  Udalski, A., 
	1999, GCN 316
\bibitem[Rees \& M\'esz\'aros 1992]{ree92} Rees, M.J. \& M\'esz\'aros, P.,
	1992, MNRAS, 258, L41
\bibitem[Sari et al. 1998]{sar98} Sari, R., Piran, T. \& Narayan, R.,
	1998, ApJ, 497, L17
\bibitem[Schlegel et al. 1998]{schle98} Schlegel, D. J., Finkbeiner, D.P. \& Davis, M.,
	1998, ApJ, 500, 525
\bibitem[Serkowski et al. 1975]{ser75} Serkowski, K., Mathewson, D.L. \& Ford, V.L.,
	1975, ApJ, 196, 261
\bibitem[Stanek et al. 1999a]{sta99a} Stanek, K.Z., Garnavich, P.M., Kaluzny, J., Pych, W. 
	\& Thompson, I., 1999b, GCN 318
\bibitem[Stanek et al. 1999b]{sta99} Stanek, K.Z., Garnavich, P.M., Kaluzny, J., 
et al. 1999, astro-ph/990534	
\bibitem[Stetson 1987]{ste87} Stetson, P. B., 1987, PASP, 99, 191
\bibitem[Vietri 1997]{vie97} Vietri, M., 1997, ApJ, 478, L9
\bibitem[Vreeswijk et al. 1999a]{vre99} Vreeswijk, P. M., Galama, T.J., 
	Rol, E. et al., 1999a, GCN 310
\bibitem[Vreeswijk et al. 1999b]{vre99b} Vreeswijk, P. M., Galama, T.J., Rol, E. et al., 
   	 1999b, GCN 324
\bibitem[Waxman 1997]{wax97} Waxman, E., 1997, ApJ, 485, L5


\end{thebibliography}
\end{document}